\documentclass[twocolumn,aps,prl,showpacs,superscriptaddress,floatfix]{revtex4}

\usepackage{color}
\usepackage{graphicx}
\usepackage{multirow}
\usepackage{amsmath,bm}

\usepackage[displaymath]{lineno}

\newcommand {\snn}	{\sqrt{s_{_{\rm NN}}}}

\newcommand {\Nch}	{N_{\rm ch}}

\newcommand {\Npart}	{N_{\rm part}}
\newcommand {\Nbin}	{N_{\rm bin}}

\newcommand {\Zr}	{$^{96}$Zr}
\newcommand {\Ru}	{$^{96}$Ru}
\newcommand {\RuRu}	{$^{96}_{44}$Ru+$^{96}_{44}$Ru}
\newcommand {\ZrZr}	{$^{96}_{40}$Zr+$^{96}_{40}$Zr}
\newcommand {\Pb}	{$^{208}$Pb}

\newcommand {\rnp}	{$\Delta r_{\rm np}$}

\newcommand {\hijing}	{Hijing}
\newcommand {\urqmd}     {UrQMD}
\newcommand {\ampt}	{AMPT}
\newcommand {\amptdef}	{AMPT-def}
\newcommand {\amptsm}	{AMPT-sm}

\newcommand {\mean}[1]	{\langle #1\rangle}

\begin{document}
\title{Probing the neutron skin with ultrarelativistic isobaric collisions}
\author{Hanlin Li}
\affiliation{College of Science, Wuhan University of Science and Technology, Wuhan, Hubei 430065, China}
\author{Hao-jie Xu\footnote{%
	Corresponding author: haojiexu@zjhu.edu.cn}}
\affiliation{School of Science, Huzhou University, Huzhou, Zhejiang 313000, China}
\author{Ying Zhou}
\affiliation{School of Physics and Astronomy and Shanghai Key Laboratory for Particle Physics and Cosmology, Shanghai Jiao Tong University, Shanghai 200240, China}
\author{Xiaobao Wang}
\affiliation{School of Science, Huzhou University, Huzhou, Zhejiang 313000, China}
\author{Jie Zhao}
\affiliation{Department of Physics and Astronomy, Purdue University, West Lafayette, Indiana 47907, USA}
\author{Lie-Wen Chen\footnote{%
	Corresponding author: lwchen@sjtu.edu.cn}}
\affiliation{School of Physics and Astronomy and Shanghai Key Laboratory for Particle Physics and Cosmology, Shanghai Jiao Tong University, Shanghai 200240, China}
\author{Fuqiang Wang\footnote{%
	Corresponding author: fqwang@zjhu.edu.cn}}
\affiliation{School of Science, Huzhou University, Huzhou, Zhejiang 313000, China}
\affiliation{Department of Physics and Astronomy, Purdue University, West Lafayette, Indiana 47907, USA}

\date{\today}

\begin{abstract}
Particle production in ultrarelativistic heavy ion collisions depends on the details of the nucleon density distributions in the colliding nuclei. We demonstrate that the charged hadron multiplicity distributions in isobaric collisions at ultrarelativistic energies provide a novel approach to determine the poorly known neutron density distributions and thus the neutron skin thickness in finite nuclei, which can in turn put stringent constraints on the nuclear symmetry energy.
\end{abstract}


\maketitle

{\em Introduction.}
Nuclei are bound states of protons and neutrons by the overall attractive nuclear force. The nuclear force is short ranged,
and is surpassed by Coulomb repulsion among protons at long distances. This is compensated by more neutrons to keep heavy nuclei bound. With more neutrons comes the penalty symmetry energy associated with the asymmetry between the proton and neutron numbers. The symmetry energy influences the proton and neutron density distributions, and in particular, the neutron skin thickness in nuclei (difference between the rms radii of the neutron and proton distributions, $\Delta r_{\rm np}\equiv r_{n}-r_{p}$)~\cite{Brown:2000pd}. The symmetry energy and its density dependence are crucial to our understanding of the masses and drip lines of neutron-rich nuclei and the equation of state (EOS) of nuclear and neutron star matter~\cite{Horowitz:2000xj,Steiner:2004fi,Li:2008gp,Wang:2014mra,Chen:2010qx,Gandolfi:2015jma,Zhang:2018bwq,Ozel:2016oaf,Baldo:2016jhp}.

Measurements of the neutron density and the \rnp, complemented by state-of-the-art theoretical calculations~\cite{Bartel:1982ed,Machleidt:1989tm,AlexBrown:1998zz,Furnstahl:2001un}, can yield valuable information on the symmetry energy~\cite{Chen:2005ti,RocaMaza:2011pm,Tsang:2012se,Horowitz:2014bja}.
Exact knowledge of nucleon density distributions is also crucial to new physics searches beyond the standard model~\cite{Huang:2019ene}.
Because protons are charged, its density distributions are well measured by electrons scattering off nuclei~\cite{Frois:1987hk,Lapikas:1003zz}.
The density distributions are not as well measured~\cite{Tsang:2012se}; for example, the \rnp\ measurements of the benchmark \Pb\ nucleus 
fall in the range of $0.15\mbox{-}0.22$ fm with a typical precision of $20\mbox{-}50\%$~\cite{ RocaMaza:2011pm,Tsang:2012se,Tarbert:2013jze}.
One limitation is the inevitable uncertainties in modeling the strong interaction of the reaction mechanisms~\cite{Ray:1992fj}. 
Promising ways to measure neutron densities are through electroweak parity-violating scattering processes with electrons~\cite{Donnelly:1989qs,Horowitz:1999fk} and neutrinos~\cite{Akimov:2017ade}, 
exploiting the large weak charge of the neutron compared to the diminishing one of the proton. 
Such measurements, although cleaner to interpret, require large luminosities~\cite{RocaMaza:2011pm,Horowitz:2013wha}. 
The current measurement by PREX (The Lead Radius Experiment) on the \Pb\ \rnp\ is $0.33_{-0.18}^{+0.16}$ fm~\cite{Abrahamyan:2012gp}.

The symmetry energy affects observables in low to intermediate energy heavy ion collisions, 
such as the isospin diffusion~\cite{Chen:2004si,Tsang:2008fd}, the neutron-proton flow difference~\cite{Li:2000bj}, the isospin dependent pion production~\cite{Li:2002qx}, and light cluster formation~\cite{Chen:2003qj}. 
Heavy ion collisions at relativistic energies are generally considered insensitive to nuclear structures and the symmetry energy.
Recent studies of isobaric \RuRu~ and \ZrZr~ collisions at nucleon-nucleon center-of-mass energy of $\snn=200$ GeV indicate, however, that nuclear density distributions have a noticeable
effect on the total charged hadron multiplicity ($\Nch$)~\cite{ Xu:2017zcn,Li:2018oec}.
Since $\Nch$ can be measured precisely, we demonstrate in this work that the $\Nch$ distributions in isobaric collisions can be used to determine the \rnp\ (and hence the symmetry energy) to a precision that may be comparable to or even exceed those achieved by traditional low energy nuclear experiments. 

{\em The symmetry energy and the neutron skin.}
The symmetry energy encodes the energy related to neutron-proton asymmetry in
nuclear matter EOS. 
It is conventionally defined in the binding energy per nucleon, $E(\rho,\delta)=E_0(\rho)+E_{\mathrm{sym}}(\rho)\delta^2 +{\cal O}(\delta^4)$,
where $\rho=\rho_{n}+\rho_{p}$ is the nucleon number density
and $\delta=(\rho_{n}-\rho_{p})/\rho$ is the isospin asymmetry
with $\rho_{p}$ ($\rho_{n}$) denoting
the proton (neutron) density~\cite{Li:2008gp}.
The symmetry energy can be obtained as
$E_{\rm sym}(\rho)=\left.\frac{1}{2}\frac{\partial^{2}E(\rho,\delta)}{\partial\delta^{2}}\right|_{\delta=0}.$
It can be expanded at $\rho_r$ in
$\chi_r=(\rho-\rho_r)/3\rho_{r}$ as
$E_{\rm sym}(\rho) = E_{\rm sym}(\rho_r) + L(\rho_r) \chi_r + \mathcal{O}(\chi_r^2)$,
where $L (\rho_r) = \left. 3\rho_r\frac{dE_{\rm sym}(\rho)}{d\rho}\right|_{\rho=\rho_r}$ 
is the density slope parameter~\cite{Li:2008gp}. 
Especially, for $\rho_r = \rho_0 \approx 0.16$~fm$^{-3}$ (the nuclear saturation density), 
one has $L \equiv L (\rho_0)$ which characterizes
the density dependence of the $E_{\rm sym}(\rho)$ around $\rho_0$.
In addition, at a subsaturation cross density 
$\rho_c = 0.11\rho_0/0.16 \approx 0.11$~fm$^{-3}$, roughly corresponding to 
the average density of finite nuclei, 
the $E_{\rm sym}(\rho_c)=26.65 \pm 0.20$~MeV~\cite{Zhang:2013wna} is precisely obtained 
from nuclear binding energies~\cite{Zhang:2013wna}. 
At this $\rho_c$, 
a strong constraint $L(\rho_c) = 47.3 \pm 7.8$~MeV is obtained from
the electric dipole polarizibility data of $^{208}$Pb~\cite{Zhang:2014yfa}. 
Generally, it is found that  the $L(\rho_c)$ displays a particularly
strong positive correlation with the \rnp\ of heavy nuclei.

In the present work, we use two different nuclear energy density functionals to 
describe nuclear matter EOS and the properties of finite nuclei, namely,
the standard Skyrme-Hartree-Fock (SHF) model (see, e.g., Ref.~\cite{Chabanat:1997qh}) and the extended SHF (eSHF) model~\cite{Chamel:2009yx,Zhang:2015vaa}.
These two models have been very successful in describing the structures of finite nuclei,
especially global properties such as binding energies and charge radii.
Compared to SHF,
the eSHF contains additional momentum and density-dependent two-body forces
to effectively simulate the momentum dependence of the three-body forces~\cite{Zhang:2015vaa}.
Fitting to data using the strategy in Ref.~\cite{Zhou:2019omw}, we obtain
an interaction parameter set (denoted as Lc47) within eSHF by fixing $L(\rho_c) = 47.3$ MeV~\cite{Zhang:2014yfa} 
with $E_{\rm sym}(\rho_c) = 26.65$~MeV~\cite{Zhang:2013wna}.
We also construct two more parameter sets (Lc20 and Lc70) with $L(\rho_c) = 20$~MeV and $70$~MeV, respectively,
keeping the same $E_{\rm sym}(\rho_c)$~\cite{Zhang:2013wna}, to explore the effects of the symmetry energy (and neutron skin) variations.
For the SHF calculations, we use the well-known interaction set SLy4~\cite{Chabanat:1997un,Wang:2016rqh}.

\begin{table}
	\caption{The rms radii (in fm) for neutron ($r_{n}$) and proton ($r_{p}$) distributions and 
    the neutron skin thickness ($\Delta r_{\rm np}\equiv r_{n}-r_{p}$) of \Ru\ and \Zr, and the symmetry energy slope parameters
	$L(\rho_{c})$ and $L$  (in Mev) from eSHF (parameter sets Lc20, Lc47, Lc70) and SHF (SLy4) calculations. The \rnp\ values for \Pb\ are also listed 
    for comparison.}
  \label{tab:R}
	\begin{tabular}{c|c|c|ccc|ccc|c}
	  \hline
	  \hline
		&   &        & \multicolumn{3}{c|}{\Zr} & \multicolumn{3}{c|}{\Ru} & \Pb \\
		& $L(\rho_c)$  & $L$ & $r_{n}$ & $r_{p}$& \rnp & $r_{n}$& $r_{p}$ & \rnp & \rnp \\ \hline
		Lc20 & 20   &13.1 & 4.386 & 4.27  & 0.115 & 4.327 & 4.316 & 0.011 & 0.109 \\ \hline
		Lc47 & 47.3 &55.7 & 4.449 & 4.267 & 0.183 & 4.360 & 4.319 & 0.042 & 0.190 \\ \hline
		Lc70 & 70   &90.0 & 4.494 & 4.262 & 0.232 & 4.385 & 4.32  & 0.066 & 0.264 \\ \hline
		SLy4 & 42.7 &46.0 & 4.432 & 4.271 & 0.161 & 4.356 & 4.327 & 0.030 & 0.160\\ \hline
  \end{tabular}
\end{table}

\begin{figure} 
	\includegraphics[scale=0.40]{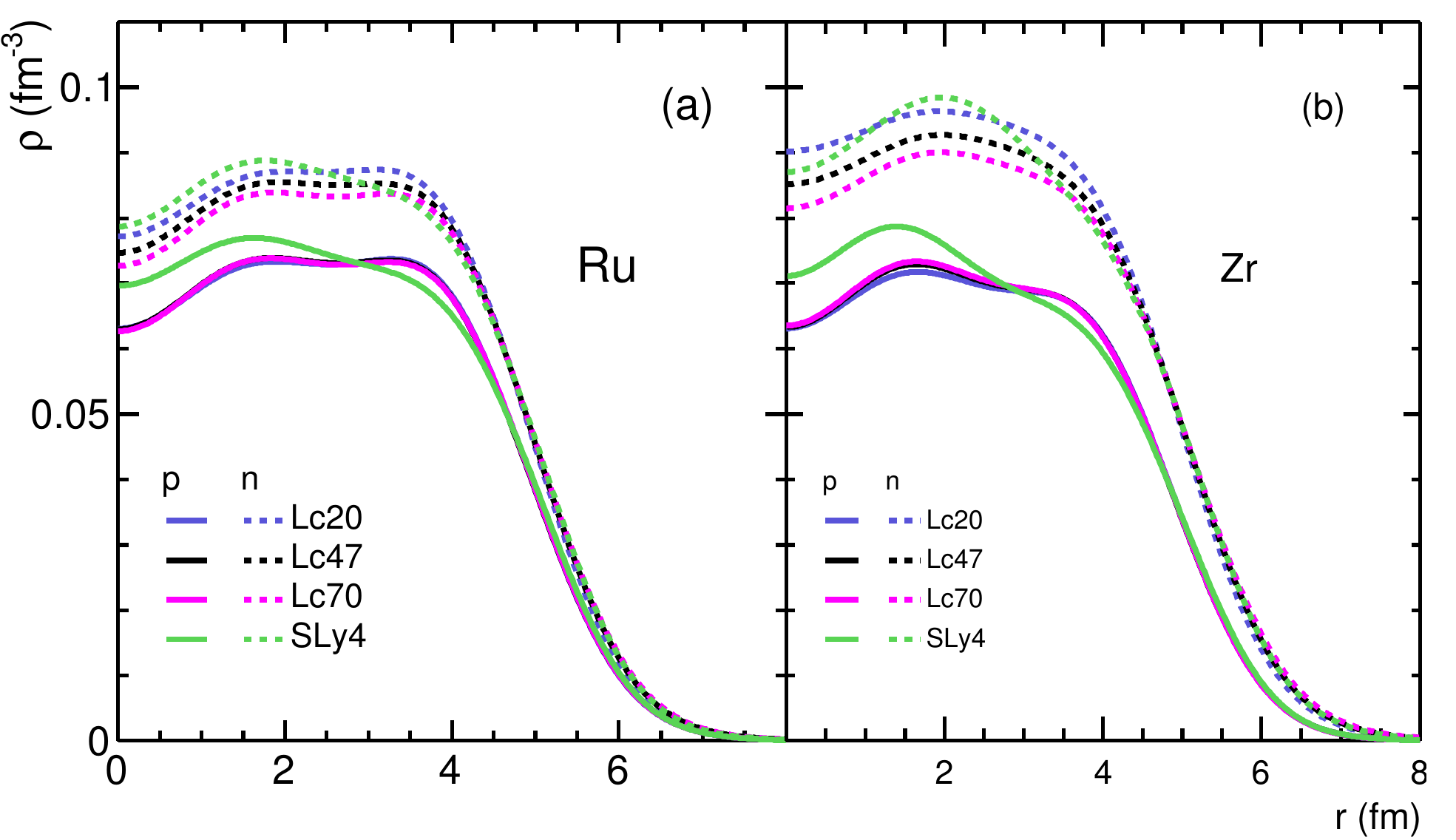}
	\caption{(Color online).
	Proton and neutron density distributions of (a) \Ru~ and (b) \Zr~ nuclei
	from eSHF (parameter sets Lc20, Lc47, Lc70) and SHF (SLy4).
	\label{fig:densities}} 
\end{figure}

Table~\ref{tab:R} lists the nuclear radii of {\Zr } and {\Ru }, assuming spherical symmetry, from the eSHF calculations
using Lc20, Lc47 and Lc70, and the SHF calculation with SLy4, together with the $L(\rho_c)$ and $L$ parameters. 
It is seen that the four interactions give similar proton rms radius $r_p$
for {\Zr } and {\Ru } since they are experimentally well constrained,
but the neutron radius $r_n$ increases with $L(\rho_c)$ and $L$,
leading to a positive correlation between \rnp\ and $L(\rho_c)$ (and $L$) as expected.
The \rnp\ of \Pb\ from our calculations are also listed in Tab.~\ref{tab:R}.
We note that those values essentially cover the current uncertainty in the \Pb\ measurements.
Shown in Fig.~\ref{fig:densities} are the corresponding nucleon density distributions of \Ru\ and \Zr.
In the following, we use these density distributions in heavy ion collision models
to examine the effects on $\Nch$.

{\em Heavy ion collision models.}
We use four typical, well developed, commonly used models for relativistic heavy ion collisions.
The \hijing\ (Heavy ion jet interaction generator, v1.411)~\cite{Wang:1991hta,Wang:1996yf} is an event generator of heavy ion collisions using binary nucleon-nucleon (NN) collisions based on the 
Glauber theory,  incorporating nuclear shadowing effect and partonic energy loss in medium. 
Each NN collision is described by multiple mini-jet production inspired by perturbative Quantum Chromodynamics (QCD),
with the LUND~\cite{Andersson:1983ia} string fragmentation.
The  \ampt\ (A Multi-Phase Transport) is a dynamical model~\cite{Zhang:1999bd}. Its default version (\amptdef, v1.26) uses \hijing\ but subjects the mini-jet partons to partonic scatterings 
via ZPC (Zhang's parton cascade)~\cite{Zhang:1997ej} and, after fragmentation, hadronic scatterings via ART (a relativistic transport)~\cite{Li:1995pra}. 
The string melting version of \ampt\ (\amptsm, v2.26)~\cite{Lin:2001zk} converts all hadrons from \hijing\ to partons 
to undergo partonic scatterings, and uses a simple coalescence for hadronization, followed by hadronic rescatterings. 
The \urqmd\ (Ultra relativistic Quantum Molecular Dynamics, v3.4)~\cite{Bass:1998ca,Bleicher:1999xi} is a microscopic transport model with covariant propagation of hadrons on classical trajectories,
combined with stochastic binary scatterings, color string formation and resonance decays.
Except for the input neutron and proton density distributions, all parameters are set to default. 
About $3\times10^{7}$ events within the impact parameter ($b$) range $[0,20]$ fm are simulated in each model for each set of the nuclear densities
for Ru+Ru and Zr+Zr collisions at $\snn=200$ GeV.

\begin{figure} 
	\includegraphics[scale=0.2]{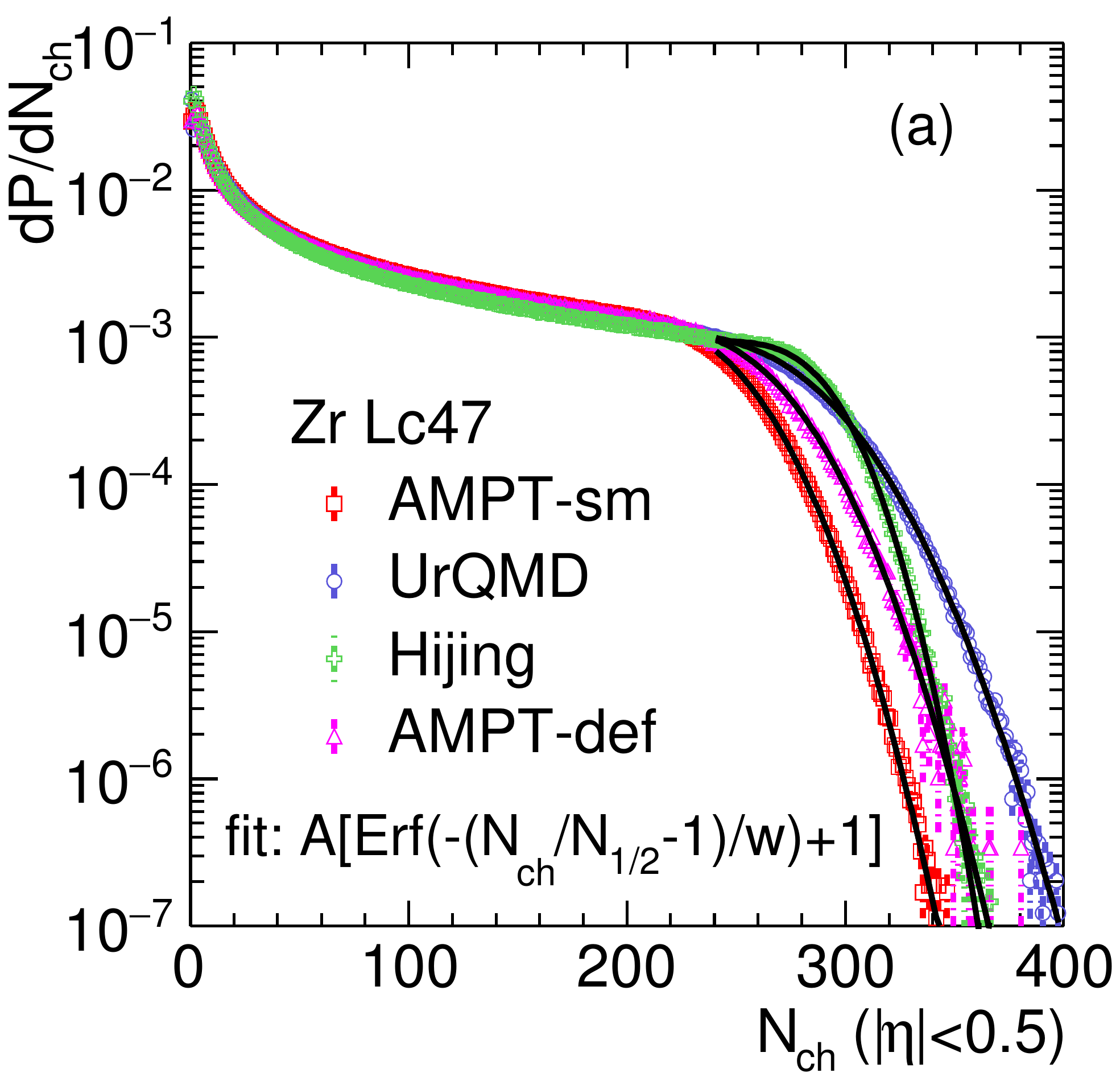}\includegraphics[scale=0.2]{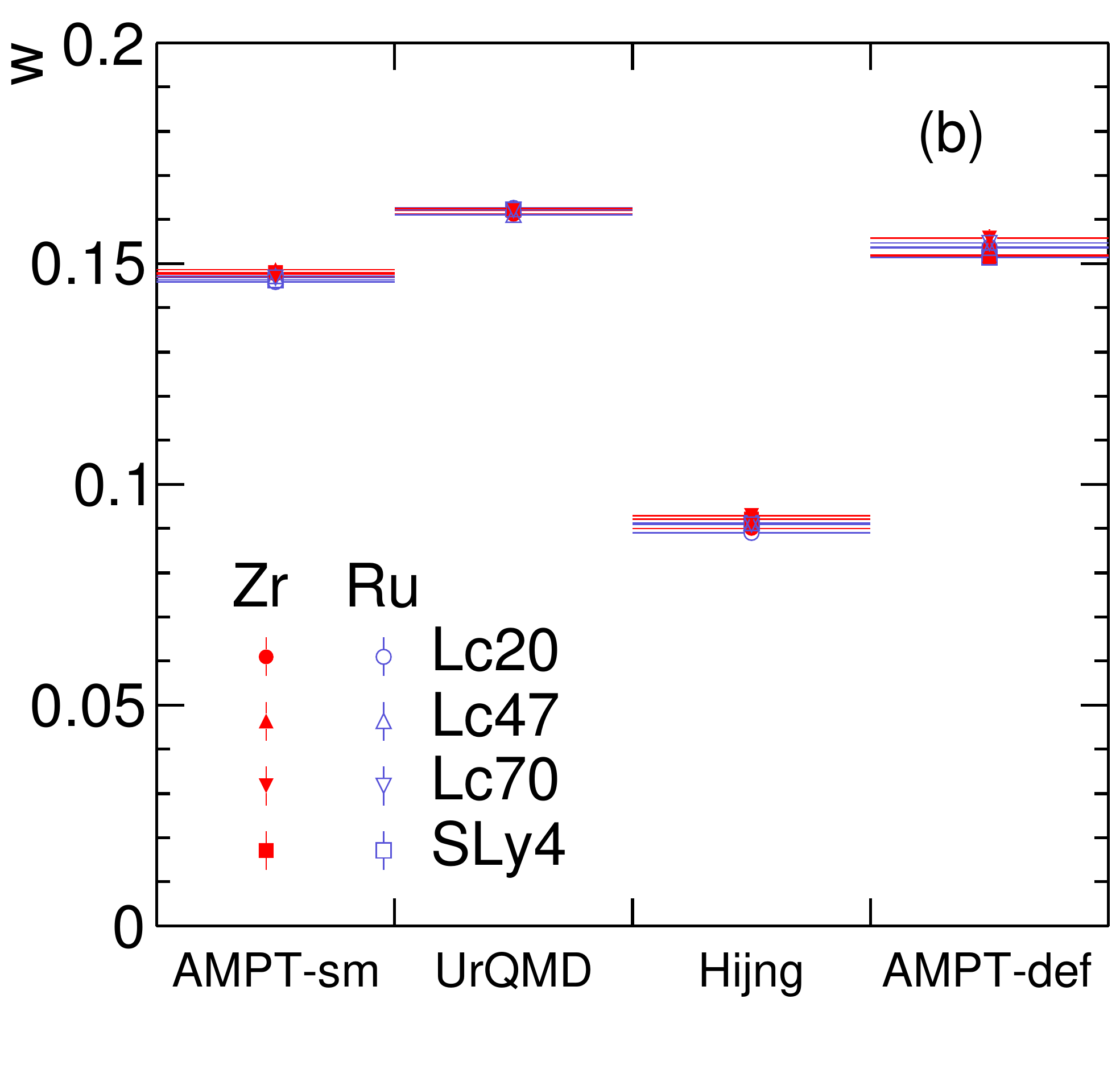}
	\caption{(Color online).
	(a) Charged hadron multiplicity ($\Nch$) distributions from \amptsm, \urqmd, \hijing, and \amptdef\ for density set Lc47. The results for the other density sets are similar. (b) The relative widths, $w$, of the $\Nch$ tails for four density sets in four models. 
	\label{fig:distribution}} 
\end{figure}

{\em Model results and discussions.}
Charged hadrons are counted with transverse momentum $p_{T}>0.2$ GeV/c and pseudo-rapidity $|\eta|<0.5$.
Figure ~\ref{fig:distribution}(a) shows the $\Nch$ distributions in Zr+Zr collisions calculated by the four models using the nuclear density set Lc47.
The distributions are similar except at large $\Nch$. The absolute $\Nch$ values are subject to large model dependence
because particle production in heavy ion collisions is generally hard to model precisely.
The shape of the $\Nch$ distribution is, on the other hand, more robust. It is primarily determined by the interaction cross-section as a function of $b$.
While the tail fall-off shapes are similar among \amptsm, \amptdef, and \urqmd, that of \hijing\ is distinct. To quantify the shape, we fit the tail distributions by 
\begin{linenomath}
\begin{equation}
	dP/d\Nch \propto -{\rm Erf}(-(\Nch/N_{1/2}-1)/w) + 1,
\end{equation}
\end{linenomath}
where $N_{1/2}$ is the $\Nch$ value at half height and $w$ is the width of the tail relative to  $N_{1/2}$. 
The fitted curves are superimposed in Fig.~\ref{fig:distribution}(a). Figure~\ref{fig:distribution}(b) depicts the fit $w$ values.
The \hijing\  model has a factor of $\sim2$ narrower tail than the other three transport models which are similar.
This feature can be readily used to distinguish models once data are available, though not the main goal of this work. 

\begin{figure} 
	\includegraphics[scale=0.28]{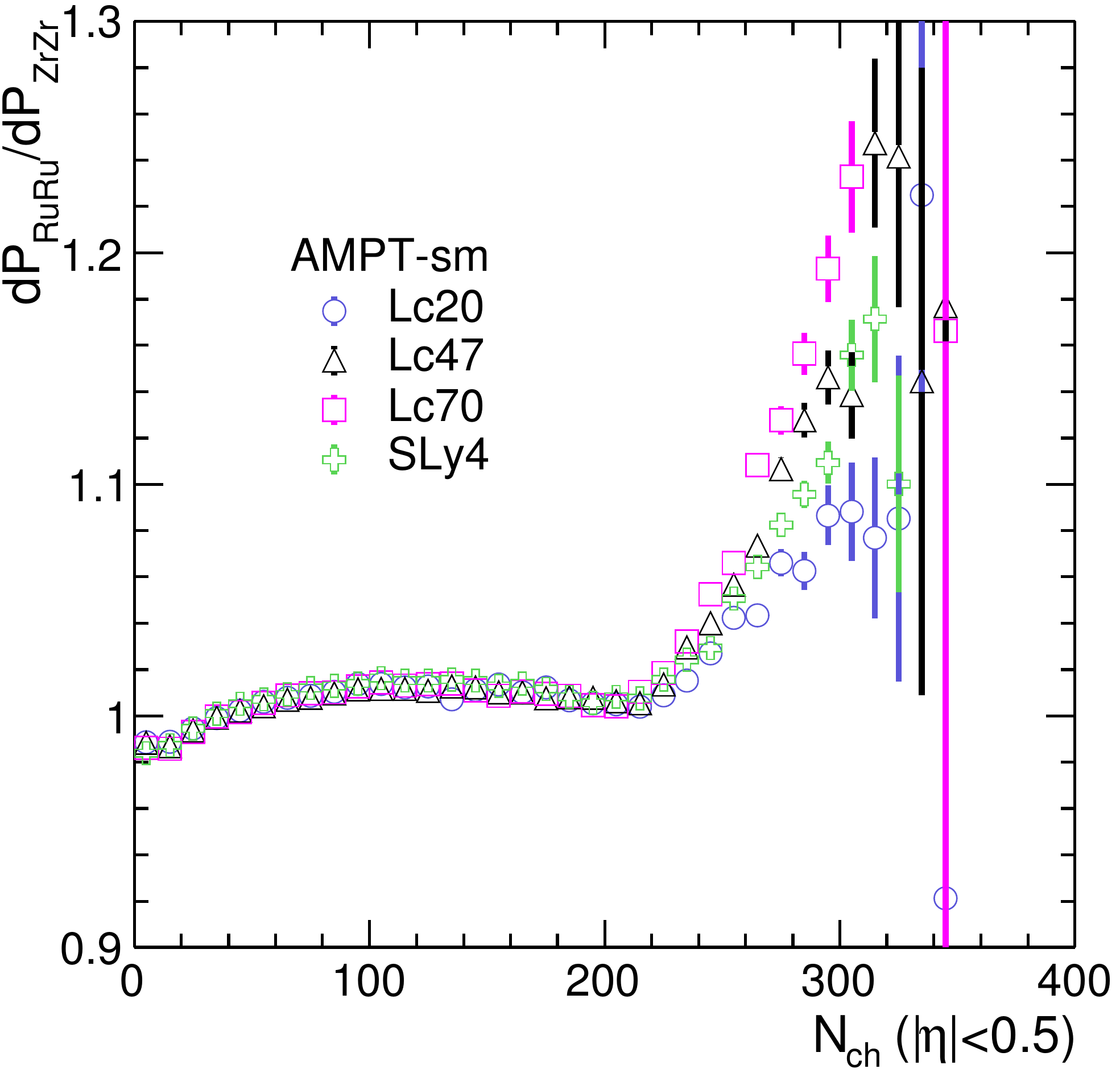}
	\caption{(Color online). Ratio of the $\Nch$ distribution in Ru+Ru to that in Zr+Zr for various densities in \amptsm. The other models are similar.
	\label{fig:RatioDis}} 
\end{figure}

The main goal of this work is to identify which density set would best describe data and hence 
to determine the neutron skin thickness and the symmetry energy. 
In a given model, at a given $b$, the numbers of participants ($\Npart$) and binary nucleon-nucleon collisions ($\Nbin$) slightly differ for different nuclear densities.
Since $\Nch$ is generally considered to depend on $\Npart$ and perhaps $\Nbin$, those differences could produce an effect on $\Nch$.
The effect is understandably small, hardly observable in a plot of the $\Nch$ distributions themselves, but can be magnified by the ratio of the $\Nch$ distribution in Ru+Ru to that in Zr+Zr~\cite{Li:2018oec}.
These ratios using the four sets of densities, in \amptsm\ as an example, are shown in Fig.~\ref{fig:RatioDis}. The splittings of the $\Nch$ tails are clear. 

The ratios in Fig.~\ref{fig:RatioDis} are illustrative to highlight the differences but are cumbersome to quantify. As seen from Fig.~\ref{fig:distribution}(b), the tail widths are equal among the densities in a given model, so the splittings are mostly due to the slight shifts in $N_{1/2}$, or differences in the average $\Nch$ values. 
The $N_{1/2}$ value is sensitive to the chosen fit range. 
We thus use the relative $\mean{\Nch}$ difference between Ru+Ru and Zr+Zr,
\begin{linenomath}
\begin{equation}
	R = 2\frac{\mean{\Nch}_{\rm RuRu}-\mean{\Nch}_{\rm ZrZr}}{\mean{\Nch}_{\rm RuRu}+\mean{\Nch}_{\rm ZrZr}},
\end{equation}
\end{linenomath}
to quantify the splitting of the $\Nch$ tails. Experimental measurements of $\Nch$ are affected by tracking inefficiency,
usually multiplicity dependent~\cite{Abelev:2008ab}. While this effect is mostly canceled in $R$, it is better to use only central collisions, say top $5\%$, where the tracking efficiency is constant to a good degree. To experimentally determine the centrality percentage, the peripheral collisions that are not recorded because of online trigger inefficiency should be taken into account. 
This trigger inefficiency can be experimentally corrected, and is equal between the isobar systems as a function of $\Nch$.
Even without correction, taking a conservative trigger efficiency of $95\%$~\cite{Abelev:2008ab}, the less than $2\%$ difference in very peripheral collisions~\cite{Li:2018oec} would yield only $10^{-3}$ mismatch in the centralities between the two systems.
This would give a negligible uncertainty on $R$ on the order of $2\times10^{-4}$.
In short, since $R$ is a relative measure between Ru+Ru and Zr+Zr collisions, much of the experimental effects cancel. 

\begin{figure} 
	\includegraphics[scale=0.35]{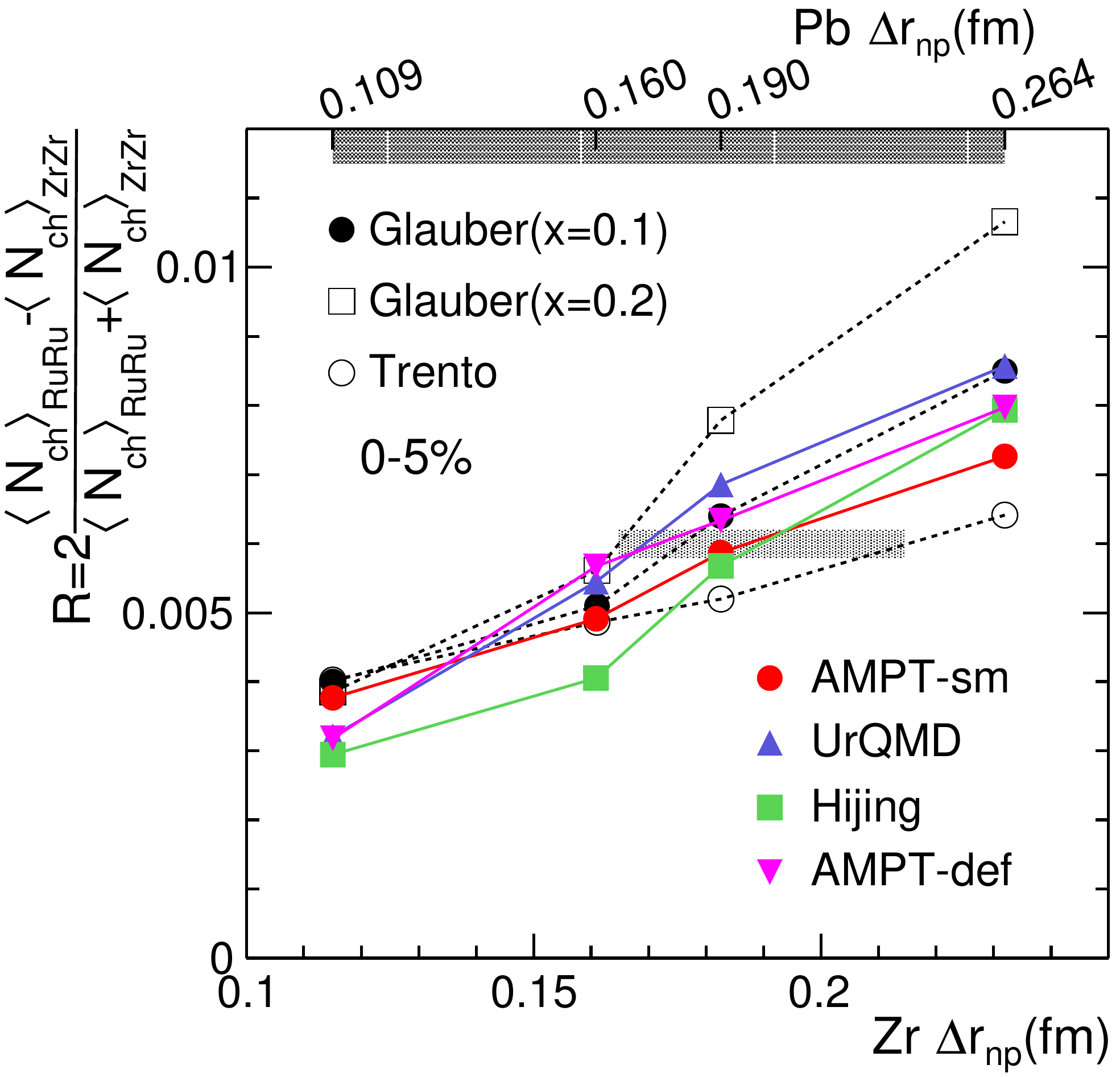}
	\caption{(Color online). The relative $\mean{\Nch}$ ratio $R$ as a function of the Zr neutron skin thickness. The four sets of data points in order from left to right are from Lc20,  SLy4, Lc47, Lc70 densities. 
	The results from \amptsm, \urqmd, \hijing, and \amptdef\ are connected by solid lines.
	The results from Glauber and Trento models are connected by dashed lines.
	\label{fig:RatioNch}} 
\end{figure}

The $R$ in each model must depend on how much the Ru and Zr nuclear density distributions differ, which can be characterized by \rnp\ of the Zr (or Ru) nucleus. 
We therefore plot in Fig.~\ref{fig:RatioNch} the $R$ in the top $5\%$ centrality against \rnp\ of the Zr nucleus from the eSHF (SHF) calculations with Lc20, Lc47 and Lc70 (SLy4). It is found that $R$ monotonically increases with \rnp. This is because, with increasing \rnp, the difference between Ru and Zr densities increases. This results in an increasing difference in $\Nch$ between Ru+Ru and Zr+Zr collisions. 

Figure~\ref{fig:RatioNch} further shows that the value of $R$ has a relatively weak model dependence. 
This includes even Hijing whose tail distribution is significantly narrower than the other models (cf. Fig.~\ref{fig:distribution}),
which can already be discriminated by data measurements as aforementioned. 
In what follows, we further decipher what $R$ entails by using non-dynamical but intuitive geometrical models, namely, the Glauber and Trento models. 

In a Glauber model~\cite{Kharzeev:2000ph,Kharzeev:2001yq}, it is postulated that $\Nch\propto [(1-x)\Npart/2+x\Nbin]$,
where $x$ is the so-called hard-component fraction.
The $\Npart$ and $\Nbin$ differ between the two isobar collisions, in a manner sensitive to the nuclear density parameters. 
This yields a non-vanishing $R$ dependence of \rnp, as shown by the dynamical model results. In addition,
the different contribution weights from $\Npart$ and $\Nbin$ to $\Nch$, characterized by the $x$ parameter, also affect the $R$.
The four models giving similar $R$ as function of \rnp\ may be indicative of their similar $x$ values. This is not surprising because all these models can 
approximately describe the centrality dependence of $\Nch$ observed by the PHOBOS experiment, which suggested an $x$ value of $0.1$~\cite{Back:2002uc}.
In order to investigate the sensitivity of $R$ to the $x$ value, we show in Fig.~\ref{fig:RatioNch} results from the Glauber model~\cite{Alver:2006wh,Miller:2007ri,Xu:2014ada} with $x=0.1$ and a significantly larger value of $x=0.2$ (which can be taken as an extreme).
Indeed the $x=0.1$ results fall within the range of the four dynamical models.
The $x=0.2$ results have a stronger sensitivity to \rnp, 
indicating that the $\Nbin$ is more sensitive to the nuclear density distribution.

On the other hand, the two-component particle production model has drawbacks. The centrality dependences of $\Nch$ at
$\snn=200$ GeV and $2.76$ TeV have essentially the same shape~\cite{Aamodt:2010cz}, 
whereas the hard-scattering cross-section, hence the $x$ value, should vary with energy.
The PHENIX experiment found that a simple wounded quark model can successfully describe 
the centrality dependence of $\Nch$ in Au+Au collisions over $7.7$--$200$ GeV~\cite{Adare:2015bua}  
and in several small systems at 200 GeV~\cite{Adare:2018toe}. 
Recent azimuthal anisotropy data by STAR in ultracentral U+U collisions cannot be described by the Glauber approach~\cite{Wang:2014qxa} and the Trento model was proposed~\cite{Moreland:2014oya}.
In the Trento model, particle production is only related to the reduced thickness, $\Nch\propto T_{R}(p;T_{A},T_{B})\equiv[(T_{A}^{p} + T_{B}^{p})/2]^{1/p}$~\cite{Bernhard:2016tnd,Moreland:2014oya}.
We use the parameter $p=0$ (i.e., $\Nch\propto\sqrt{T_AT_B}$), a gamma fluctuation parameter $k=1.4$, and a Gaussian nucleon size of $0.6$ fm, which were found to well describe the multiplicity data in heavy ion collisions~\cite{Bernhard:2016tnd,Moreland:2014oya}. 
The $R$ calculated by the Trento model is shown in Fig.~\ref{fig:RatioNch}. 
The \rnp\ dependence is weaker because only $\Npart$ contributes to the $\Nch$.
Considering all model results, the overall spread in $R$ is 
wider at larger \rnp. This is because the nuclear density difference between the Ru and Zr nuclei
increases with increasing \rnp, so does the model dependence.

Experimentally, the $\Nch$ can be measured exquisitely precisely. The relative $\mean{\Nch}$ difference in central collisions is immune to many experimental uncertainties. 
Figure~\ref{fig:RatioNch} thus strongly suggests that 
the isobar data may determine \rnp\ relatively accurately. This is afforded by the rather weak dependence of the $R$ observable to
the details of QCD physics on particle production. This is in contrast to the hadronic observables in previous low-energy studies,
where strong model dependences prevent a more precise determination~\cite{Tsang:2012se,Tarbert:2013jze}. 
The current experimental range of the \Pb\ \rnp\ is indicated by the band on the top of Fig.~\ref{fig:RatioNch}. 
The \Pb\ \rnp\ calculated by the eSHF (SHF) to accommodate those low-energy measurements are written above the band.
Our results in Fig.~\ref{fig:RatioNch} indicate that with a given measurement of $R$, the precision in the derived \rnp\ of \Zr\ can be as good as $0.05$ fm covered by the four dynamical models(\amptsm, \urqmd, \hijing, \amptdef) and the two static models (Trento and Glauber),
as illustrated by the lower band (taking hypothetically $R=0.006$). 
This would be an appreciable improvement over the current constraint from \Pb.
Our results shall thus provide a significant input to help constrain the symmetry energy, bearing important implications to nuclear matter and neutron star EOS.

We have assumed spherical nuclei in our calculations. The main idea of our work is still valid with deformed nuclei. 
There are a number of promising ways to determine the nuclear deformity from heavy ion collisions~\cite{Goldschmidt:2015kpa,Giacalone:2018apa,Pang:2019aqb,Giacalone:2019pca}. 
We postpone such a study to a future work.

We focused on central collisions only, in part because of the technical advantages aforementioned.
However, one would naively expect the neutron skin effect to be the strongest in peripheral collisions.
This would be true if one could uniquely determine and sort events in $b$. 
Experimentally, the collision centrality is usually determined by final-state particle multiplicity. 
In peripheral low multiplicity events, a wide range of $b$ is mixed due to large fluctuations so that
the nuclear density difference is mostly smeared out.

We note that the final state $\Nch$  is sensitive only to the overall nucleon density at relativistic energies, and therefore indirectly sensitive to the neutron
density (and neutron skin) given that the proton density is well determined. 
In low energy nuclear reactions, isospin-sensitive observables~\cite{Xiao:2008vm,Wei:2013sfa,Hartnack:2018sih, Helenius:2016dsk} were extensively studied where
the isospin-dependent interactions are important.
Note that we have studied isobaric collisions at fairly high energy
of $\snn=200$ GeV primarily because those data have already been taken so that our method can be readily applied.

{\em Conclusions.}
We have calculated nuclear densities by energy density functional theory using several symmetry energy parameters.
We show, using four dynamical heavy ion collision models and two static models, that the charged hadron multiplicity difference
between isobar \RuRu\ and \ZrZr\ collisions has a strong sensitivity to the neutron skin and the symmetry energy, with weak model dependence.
Because the charged hadron multiplicity can be precisely measured and because the systematic uncertainties largely cancel,
our findings suggest potentially significant improvement to neutron skin and symmetry energy determination using relativistic isobar collision data collected in 2018 at the Relativistic Heavy Ion Collider.

{\em Acknowledgments.}
This work is supported in part by the National Natural Science Foundation of 
China (Grant Nos. 11905059, 11625521, U1732138, 11605054, 11505056, 11847315, 11947410),
the Major State Basic Research Development Program (973 Program) in China under Contract No. 2015CB856904,
the Natural Science Foundation of Hubei Province under Grant No. 2019CFB563,
and the U.S. Department of Energy (Grant No. DE-SC0012910).  
HX acknowledges financial support from the China Scholarship Council.

\bibliography{ref}

\begin{thebibliography}{73}
\expandafter\ifx\csname natexlab\endcsname\relax\def\natexlab#1{#1}\fi
\expandafter\ifx\csname bibnamefont\endcsname\relax
  \def\bibnamefont#1{#1}\fi
\expandafter\ifx\csname bibfnamefont\endcsname\relax
  \def\bibfnamefont#1{#1}\fi
\expandafter\ifx\csname citenamefont\endcsname\relax
  \def\citenamefont#1{#1}\fi
\expandafter\ifx\csname url\endcsname\relax
  \def\url#1{\texttt{#1}}\fi
\expandafter\ifx\csname urlprefix\endcsname\relax\def\urlprefix{URL }\fi
\providecommand{\bibinfo}[2]{#2}
\providecommand{\eprint}[2][]{\url{#2}}

\bibitem[{\citenamefont{Brown}(2000)}]{Brown:2000pd}
\bibinfo{author}{\bibfnamefont{B.~A.} \bibnamefont{Brown}},
  \bibinfo{journal}{Phys. Rev. Lett.} \textbf{\bibinfo{volume}{85}},
  \bibinfo{pages}{5296} (\bibinfo{year}{2000}).

\bibitem[{\citenamefont{Horowitz and Piekarewicz}(2001)}]{Horowitz:2000xj}
\bibinfo{author}{\bibfnamefont{C.~J.} \bibnamefont{Horowitz}} \bibnamefont{and}
  \bibinfo{author}{\bibfnamefont{J.}~\bibnamefont{Piekarewicz}},
  \bibinfo{journal}{Phys. Rev. Lett.} \textbf{\bibinfo{volume}{86}},
  \bibinfo{pages}{5647} (\bibinfo{year}{2001}), \eprint{astro-ph/0010227}.

\bibitem[{\citenamefont{Steiner et~al.}(2005)\citenamefont{Steiner, Prakash,
  Lattimer, and Ellis}}]{Steiner:2004fi}
\bibinfo{author}{\bibfnamefont{A.~W.} \bibnamefont{Steiner}},
  \bibinfo{author}{\bibfnamefont{M.}~\bibnamefont{Prakash}},
  \bibinfo{author}{\bibfnamefont{J.~M.} \bibnamefont{Lattimer}},
  \bibnamefont{and} \bibinfo{author}{\bibfnamefont{P.~J.} \bibnamefont{Ellis}},
  \bibinfo{journal}{Phys. Rept.} \textbf{\bibinfo{volume}{411}},
  \bibinfo{pages}{325} (\bibinfo{year}{2005}), \eprint{nucl-th/0410066}.

\bibitem[{\citenamefont{Li et~al.}(2008)\citenamefont{Li, Chen, and
  Ko}}]{Li:2008gp}
\bibinfo{author}{\bibfnamefont{B.-A.} \bibnamefont{Li}},
  \bibinfo{author}{\bibfnamefont{L.-W.} \bibnamefont{Chen}}, \bibnamefont{and}
  \bibinfo{author}{\bibfnamefont{C.~M.} \bibnamefont{Ko}},
  \bibinfo{journal}{Phys. Rept.} \textbf{\bibinfo{volume}{464}},
  \bibinfo{pages}{113} (\bibinfo{year}{2008}), \eprint{0804.3580}.

\bibitem[{\citenamefont{Wang and Chen}(2015)}]{Wang:2014mra}
\bibinfo{author}{\bibfnamefont{R.}~\bibnamefont{Wang}} \bibnamefont{and}
  \bibinfo{author}{\bibfnamefont{L.-W.} \bibnamefont{Chen}},
  \bibinfo{journal}{Phys. Rev.} \textbf{\bibinfo{volume}{C92}},
  \bibinfo{pages}{031303} (\bibinfo{year}{2015}), \eprint{1410.2498}.

\bibitem[{\citenamefont{Chen et~al.}(2010)\citenamefont{Chen, Ko, Li, and
  Xu}}]{Chen:2010qx}
\bibinfo{author}{\bibfnamefont{L.-W.} \bibnamefont{Chen}},
  \bibinfo{author}{\bibfnamefont{C.~M.} \bibnamefont{Ko}},
  \bibinfo{author}{\bibfnamefont{B.-A.} \bibnamefont{Li}}, \bibnamefont{and}
  \bibinfo{author}{\bibfnamefont{J.}~\bibnamefont{Xu}}, \bibinfo{journal}{Phys.
  Rev.} \textbf{\bibinfo{volume}{C82}}, \bibinfo{pages}{024321}
  (\bibinfo{year}{2010}), \eprint{1004.4672}.

\bibitem[{\citenamefont{Gandolfi et~al.}(2015)\citenamefont{Gandolfi, Gezerlis,
  and Carlson}}]{Gandolfi:2015jma}
\bibinfo{author}{\bibfnamefont{S.}~\bibnamefont{Gandolfi}},
  \bibinfo{author}{\bibfnamefont{A.}~\bibnamefont{Gezerlis}}, \bibnamefont{and}
  \bibinfo{author}{\bibfnamefont{J.}~\bibnamefont{Carlson}},
  \bibinfo{journal}{Ann. Rev. Nucl. Part. Sci.} \textbf{\bibinfo{volume}{65}},
  \bibinfo{pages}{303} (\bibinfo{year}{2015}), \eprint{1501.05675}.

\bibitem[{\citenamefont{Zhang and Li}(2019)}]{Zhang:2018bwq}
\bibinfo{author}{\bibfnamefont{N.-B.} \bibnamefont{Zhang}} \bibnamefont{and}
  \bibinfo{author}{\bibfnamefont{B.-A.} \bibnamefont{Li}},
  \bibinfo{journal}{Eur. Phys. J.} \textbf{\bibinfo{volume}{A55}},
  \bibinfo{pages}{39} (\bibinfo{year}{2019}), \eprint{1807.07698}.

\bibitem[{\citenamefont{{\"{O}}zel and Freire}(2016)}]{Ozel:2016oaf}
\bibinfo{author}{\bibfnamefont{F.}~\bibnamefont{{\"{O}}zel}} \bibnamefont{and}
  \bibinfo{author}{\bibfnamefont{P.}~\bibnamefont{Freire}},
  \bibinfo{journal}{Ann. Rev. Astron. Astrophys.}
  \textbf{\bibinfo{volume}{54}}, \bibinfo{pages}{401} (\bibinfo{year}{2016}),
  \eprint{1603.02698}.

\bibitem[{\citenamefont{Baldo and Burgio}(2016)}]{Baldo:2016jhp}
\bibinfo{author}{\bibfnamefont{M.}~\bibnamefont{Baldo}} \bibnamefont{and}
  \bibinfo{author}{\bibfnamefont{G.~F.} \bibnamefont{Burgio}},
  \bibinfo{journal}{Prog. Part. Nucl. Phys.} \textbf{\bibinfo{volume}{91}},
  \bibinfo{pages}{203} (\bibinfo{year}{2016}), \eprint{1606.08838}.

\bibitem[{\citenamefont{Bartel et~al.}(1982)\citenamefont{Bartel, Quentin,
  Brack, Guet, and Hakansson}}]{Bartel:1982ed}
\bibinfo{author}{\bibfnamefont{J.}~\bibnamefont{Bartel}},
  \bibinfo{author}{\bibfnamefont{P.}~\bibnamefont{Quentin}},
  \bibinfo{author}{\bibfnamefont{M.}~\bibnamefont{Brack}},
  \bibinfo{author}{\bibfnamefont{C.}~\bibnamefont{Guet}}, \bibnamefont{and}
  \bibinfo{author}{\bibfnamefont{H.~B.} \bibnamefont{Hakansson}},
  \bibinfo{journal}{Nucl. Phys.} \textbf{\bibinfo{volume}{A386}},
  \bibinfo{pages}{79} (\bibinfo{year}{1982}).

\bibitem[{\citenamefont{Machleidt}(1989)}]{Machleidt:1989tm}
\bibinfo{author}{\bibfnamefont{R.}~\bibnamefont{Machleidt}},
  \bibinfo{journal}{Adv. Nucl. Phys.} \textbf{\bibinfo{volume}{19}},
  \bibinfo{pages}{189} (\bibinfo{year}{1989}).

\bibitem[{\citenamefont{Alex~Brown}(1998)}]{AlexBrown:1998zz}
\bibinfo{author}{\bibfnamefont{B.}~\bibnamefont{Alex~Brown}},
  \bibinfo{journal}{Phys. Rev.} \textbf{\bibinfo{volume}{C58}},
  \bibinfo{pages}{220} (\bibinfo{year}{1998}).

\bibitem[{\citenamefont{Furnstahl}(2002)}]{Furnstahl:2001un}
\bibinfo{author}{\bibfnamefont{R.~J.} \bibnamefont{Furnstahl}},
  \bibinfo{journal}{Nucl. Phys.} \textbf{\bibinfo{volume}{A706}},
  \bibinfo{pages}{85} (\bibinfo{year}{2002}), \eprint{nucl-th/0112085}.

\bibitem[{\citenamefont{Chen et~al.}(2005{\natexlab{a}})\citenamefont{Chen, Ko,
  and Li}}]{Chen:2005ti}
\bibinfo{author}{\bibfnamefont{L.-W.} \bibnamefont{Chen}},
  \bibinfo{author}{\bibfnamefont{C.~M.} \bibnamefont{Ko}}, \bibnamefont{and}
  \bibinfo{author}{\bibfnamefont{B.-A.} \bibnamefont{Li}},
  \bibinfo{journal}{Phys. Rev.} \textbf{\bibinfo{volume}{C72}},
  \bibinfo{pages}{064309} (\bibinfo{year}{2005}{\natexlab{a}}),
  \eprint{nucl-th/0509009}.

\bibitem[{\citenamefont{Roca-Maza et~al.}(2011)\citenamefont{Roca-Maza,
  Centelles, Vinas, and Warda}}]{RocaMaza:2011pm}
\bibinfo{author}{\bibfnamefont{X.}~\bibnamefont{Roca-Maza}},
  \bibinfo{author}{\bibfnamefont{M.}~\bibnamefont{Centelles}},
  \bibinfo{author}{\bibfnamefont{X.}~\bibnamefont{Vinas}}, \bibnamefont{and}
  \bibinfo{author}{\bibfnamefont{M.}~\bibnamefont{Warda}},
  \bibinfo{journal}{Phys. Rev. Lett.} \textbf{\bibinfo{volume}{106}},
  \bibinfo{pages}{252501} (\bibinfo{year}{2011}), \eprint{1103.1762}.

\bibitem[{\citenamefont{Tsang et~al.}(2012)}]{Tsang:2012se}
\bibinfo{author}{\bibfnamefont{M.~B.} \bibnamefont{Tsang}}
  \bibnamefont{et~al.}, \bibinfo{journal}{Phys. Rev.}
  \textbf{\bibinfo{volume}{C86}}, \bibinfo{pages}{015803}
  (\bibinfo{year}{2012}), \eprint{1204.0466}.

\bibitem[{\citenamefont{Horowitz
  et~al.}(2014{\natexlab{a}})\citenamefont{Horowitz, Brown, Kim, Lynch,
  Michaels, Ono, Piekarewicz, Tsang, and Wolter}}]{Horowitz:2014bja}
\bibinfo{author}{\bibfnamefont{C.~J.} \bibnamefont{Horowitz}},
  \bibinfo{author}{\bibfnamefont{E.~F.} \bibnamefont{Brown}},
  \bibinfo{author}{\bibfnamefont{Y.}~\bibnamefont{Kim}},
  \bibinfo{author}{\bibfnamefont{W.~G.} \bibnamefont{Lynch}},
  \bibinfo{author}{\bibfnamefont{R.}~\bibnamefont{Michaels}},
  \bibinfo{author}{\bibfnamefont{A.}~\bibnamefont{Ono}},
  \bibinfo{author}{\bibfnamefont{J.}~\bibnamefont{Piekarewicz}},
  \bibinfo{author}{\bibfnamefont{M.~B.} \bibnamefont{Tsang}}, \bibnamefont{and}
  \bibinfo{author}{\bibfnamefont{H.~H.} \bibnamefont{Wolter}},
  \bibinfo{journal}{J. Phys.} \textbf{\bibinfo{volume}{G41}},
  \bibinfo{pages}{093001} (\bibinfo{year}{2014}{\natexlab{a}}),
  \eprint{1401.5839}.

\bibitem[{\citenamefont{Huang and Chen}(2019)}]{Huang:2019ene}
\bibinfo{author}{\bibfnamefont{X.-R.} \bibnamefont{Huang}} \bibnamefont{and}
  \bibinfo{author}{\bibfnamefont{L.-W.} \bibnamefont{Chen}},
  \bibinfo{journal}{Phys. Rev.} \textbf{\bibinfo{volume}{D100}},
  \bibinfo{pages}{071301} (\bibinfo{year}{2019}), \eprint{1902.07625}.

\bibitem[{\citenamefont{Frois and Papanicolas}(1987)}]{Frois:1987hk}
\bibinfo{author}{\bibfnamefont{B.}~\bibnamefont{Frois}} \bibnamefont{and}
  \bibinfo{author}{\bibfnamefont{C.~N.} \bibnamefont{Papanicolas}},
  \bibinfo{journal}{Ann. Rev. Nucl. Part. Sci.} \textbf{\bibinfo{volume}{37}},
  \bibinfo{pages}{133} (\bibinfo{year}{1987}).

\bibitem[{\citenamefont{Lapikas}(1993)}]{Lapikas:1003zz}
\bibinfo{author}{\bibfnamefont{L.}~\bibnamefont{Lapikas}},
  \bibinfo{journal}{Nucl. Phys.} \textbf{\bibinfo{volume}{A553}},
  \bibinfo{pages}{297c} (\bibinfo{year}{1993}).

\bibitem[{\citenamefont{Tarbert et~al.}(2014)}]{Tarbert:2013jze}
\bibinfo{author}{\bibfnamefont{C.~M.} \bibnamefont{Tarbert}}
  \bibnamefont{et~al.}, \bibinfo{journal}{Phys. Rev. Lett.}
  \textbf{\bibinfo{volume}{112}}, \bibinfo{pages}{242502}
  (\bibinfo{year}{2014}), \eprint{1311.0168}.

\bibitem[{\citenamefont{Ray et~al.}(1992)\citenamefont{Ray, Hoffmann, and
  Coker}}]{Ray:1992fj}
\bibinfo{author}{\bibfnamefont{L.}~\bibnamefont{Ray}},
  \bibinfo{author}{\bibfnamefont{G.~W.} \bibnamefont{Hoffmann}},
  \bibnamefont{and} \bibinfo{author}{\bibfnamefont{W.~R.} \bibnamefont{Coker}},
  \bibinfo{journal}{Phys. Rept.} \textbf{\bibinfo{volume}{212}},
  \bibinfo{pages}{223} (\bibinfo{year}{1992}).

\bibitem[{\citenamefont{Donnelly et~al.}(1989)\citenamefont{Donnelly, Dubach,
  and Sick}}]{Donnelly:1989qs}
\bibinfo{author}{\bibfnamefont{T.~W.} \bibnamefont{Donnelly}},
  \bibinfo{author}{\bibfnamefont{J.}~\bibnamefont{Dubach}}, \bibnamefont{and}
  \bibinfo{author}{\bibfnamefont{I.}~\bibnamefont{Sick}},
  \bibinfo{journal}{Nucl. Phys.} \textbf{\bibinfo{volume}{A503}},
  \bibinfo{pages}{589} (\bibinfo{year}{1989}).

\bibitem[{\citenamefont{Horowitz et~al.}(2001)\citenamefont{Horowitz, Pollock,
  Souder, and Michaels}}]{Horowitz:1999fk}
\bibinfo{author}{\bibfnamefont{C.~J.} \bibnamefont{Horowitz}},
  \bibinfo{author}{\bibfnamefont{S.~J.} \bibnamefont{Pollock}},
  \bibinfo{author}{\bibfnamefont{P.~A.} \bibnamefont{Souder}},
  \bibnamefont{and} \bibinfo{author}{\bibfnamefont{R.}~\bibnamefont{Michaels}},
  \bibinfo{journal}{Phys. Rev.} \textbf{\bibinfo{volume}{C63}},
  \bibinfo{pages}{025501} (\bibinfo{year}{2001}), \eprint{nucl-th/9912038}.

\bibitem[{\citenamefont{Akimov et~al.}(2017)}]{Akimov:2017ade}
\bibinfo{author}{\bibfnamefont{D.}~\bibnamefont{Akimov}} \bibnamefont{et~al.}
  (\bibinfo{collaboration}{COHERENT}), \bibinfo{journal}{Science}
  \textbf{\bibinfo{volume}{357}}, \bibinfo{pages}{1123} (\bibinfo{year}{2017}),
  \eprint{1708.01294}.

\bibitem[{\citenamefont{Horowitz
  et~al.}(2014{\natexlab{b}})\citenamefont{Horowitz, Kumar, and
  Michaels}}]{Horowitz:2013wha}
\bibinfo{author}{\bibfnamefont{C.~J.} \bibnamefont{Horowitz}},
  \bibinfo{author}{\bibfnamefont{K.~S.} \bibnamefont{Kumar}}, \bibnamefont{and}
  \bibinfo{author}{\bibfnamefont{R.}~\bibnamefont{Michaels}},
  \bibinfo{journal}{Eur. Phys. J.} \textbf{\bibinfo{volume}{A50}},
  \bibinfo{pages}{48} (\bibinfo{year}{2014}{\natexlab{b}}), \eprint{1307.3572}.

\bibitem[{\citenamefont{Abrahamyan et~al.}(2012)}]{Abrahamyan:2012gp}
\bibinfo{author}{\bibfnamefont{S.}~\bibnamefont{Abrahamyan}}
  \bibnamefont{et~al.}, \bibinfo{journal}{Phys. Rev. Lett.}
  \textbf{\bibinfo{volume}{108}}, \bibinfo{pages}{112502}
  (\bibinfo{year}{2012}), \eprint{1201.2568}.

\bibitem[{\citenamefont{Chen et~al.}(2005{\natexlab{b}})\citenamefont{Chen, Ko,
  and Li}}]{Chen:2004si}
\bibinfo{author}{\bibfnamefont{L.-W.} \bibnamefont{Chen}},
  \bibinfo{author}{\bibfnamefont{C.~M.} \bibnamefont{Ko}}, \bibnamefont{and}
  \bibinfo{author}{\bibfnamefont{B.-A.} \bibnamefont{Li}},
  \bibinfo{journal}{Phys. Rev. Lett.} \textbf{\bibinfo{volume}{94}},
  \bibinfo{pages}{032701} (\bibinfo{year}{2005}{\natexlab{b}}),
  \eprint{nucl-th/0407032}.

\bibitem[{\citenamefont{Tsang et~al.}(2009)\citenamefont{Tsang, Zhang,
  Danielewicz, Famiano, Li, Lynch, and Steiner}}]{Tsang:2008fd}
\bibinfo{author}{\bibfnamefont{M.~B.} \bibnamefont{Tsang}},
  \bibinfo{author}{\bibfnamefont{Y.}~\bibnamefont{Zhang}},
  \bibinfo{author}{\bibfnamefont{P.}~\bibnamefont{Danielewicz}},
  \bibinfo{author}{\bibfnamefont{M.}~\bibnamefont{Famiano}},
  \bibinfo{author}{\bibfnamefont{Z.}~\bibnamefont{Li}},
  \bibinfo{author}{\bibfnamefont{W.~G.} \bibnamefont{Lynch}}, \bibnamefont{and}
  \bibinfo{author}{\bibfnamefont{A.~W.} \bibnamefont{Steiner}},
  \bibinfo{journal}{Phys. Rev. Lett.} \textbf{\bibinfo{volume}{102}},
  \bibinfo{pages}{122701} (\bibinfo{year}{2009}), \eprint{0811.3107}.

\bibitem[{\citenamefont{Li}(2000)}]{Li:2000bj}
\bibinfo{author}{\bibfnamefont{B.-A.} \bibnamefont{Li}},
  \bibinfo{journal}{Phys. Rev. Lett.} \textbf{\bibinfo{volume}{85}},
  \bibinfo{pages}{4221} (\bibinfo{year}{2000}), \eprint{nucl-th/0009069}.

\bibitem[{\citenamefont{Li}(2002)}]{Li:2002qx}
\bibinfo{author}{\bibfnamefont{B.-A.} \bibnamefont{Li}},
  \bibinfo{journal}{Phys. Rev. Lett.} \textbf{\bibinfo{volume}{88}},
  \bibinfo{pages}{192701} (\bibinfo{year}{2002}), \eprint{nucl-th/0205002}.

\bibitem[{\citenamefont{Chen et~al.}(2003)\citenamefont{Chen, Ko, and
  Li}}]{Chen:2003qj}
\bibinfo{author}{\bibfnamefont{L.-W.} \bibnamefont{Chen}},
  \bibinfo{author}{\bibfnamefont{C.~M.} \bibnamefont{Ko}}, \bibnamefont{and}
  \bibinfo{author}{\bibfnamefont{B.-A.} \bibnamefont{Li}},
  \bibinfo{journal}{Phys. Rev.} \textbf{\bibinfo{volume}{C68}},
  \bibinfo{pages}{017601} (\bibinfo{year}{2003}), \eprint{nucl-th/0302068}.

\bibitem[{\citenamefont{Xu et~al.}(2018)\citenamefont{Xu, Wang, Li, Zhao, Lin,
  Shen, and Wang}}]{Xu:2017zcn}
\bibinfo{author}{\bibfnamefont{H.-J.} \bibnamefont{Xu}},
  \bibinfo{author}{\bibfnamefont{X.}~\bibnamefont{Wang}},
  \bibinfo{author}{\bibfnamefont{H.}~\bibnamefont{Li}},
  \bibinfo{author}{\bibfnamefont{J.}~\bibnamefont{Zhao}},
  \bibinfo{author}{\bibfnamefont{Z.-W.} \bibnamefont{Lin}},
  \bibinfo{author}{\bibfnamefont{C.}~\bibnamefont{Shen}}, \bibnamefont{and}
  \bibinfo{author}{\bibfnamefont{F.}~\bibnamefont{Wang}},
  \bibinfo{journal}{Phys. Rev. Lett.} \textbf{\bibinfo{volume}{121}},
  \bibinfo{pages}{022301} (\bibinfo{year}{2018}), \eprint{1710.03086}.

\bibitem[{\citenamefont{Li et~al.}(2018)\citenamefont{Li, Xu, Zhao, Lin, Zhang,
  Wang, Shen, and Wang}}]{Li:2018oec}
\bibinfo{author}{\bibfnamefont{H.}~\bibnamefont{Li}},
  \bibinfo{author}{\bibfnamefont{H.-j.} \bibnamefont{Xu}},
  \bibinfo{author}{\bibfnamefont{J.}~\bibnamefont{Zhao}},
  \bibinfo{author}{\bibfnamefont{Z.-W.} \bibnamefont{Lin}},
  \bibinfo{author}{\bibfnamefont{H.}~\bibnamefont{Zhang}},
  \bibinfo{author}{\bibfnamefont{X.}~\bibnamefont{Wang}},
  \bibinfo{author}{\bibfnamefont{C.}~\bibnamefont{Shen}}, \bibnamefont{and}
  \bibinfo{author}{\bibfnamefont{F.}~\bibnamefont{Wang}},
  \bibinfo{journal}{Phys. Rev.} \textbf{\bibinfo{volume}{C98}},
  \bibinfo{pages}{054907} (\bibinfo{year}{2018}), \eprint{1808.06711}.

\bibitem[{\citenamefont{Zhang and Chen}(2013)}]{Zhang:2013wna}
\bibinfo{author}{\bibfnamefont{Z.}~\bibnamefont{Zhang}} \bibnamefont{and}
  \bibinfo{author}{\bibfnamefont{L.-W.} \bibnamefont{Chen}},
  \bibinfo{journal}{Phys. Lett.} \textbf{\bibinfo{volume}{B726}},
  \bibinfo{pages}{234} (\bibinfo{year}{2013}), \eprint{1302.5327}.

\bibitem[{\citenamefont{Zhang and Chen}(2014)}]{Zhang:2014yfa}
\bibinfo{author}{\bibfnamefont{Z.}~\bibnamefont{Zhang}} \bibnamefont{and}
  \bibinfo{author}{\bibfnamefont{L.-W.} \bibnamefont{Chen}},
  \bibinfo{journal}{Phys. Rev.} \textbf{\bibinfo{volume}{C90}},
  \bibinfo{pages}{064317} (\bibinfo{year}{2014}), \eprint{1407.8054}.

\bibitem[{\citenamefont{Chabanat et~al.}(1997)\citenamefont{Chabanat, Meyer,
  Bonche, Schaeffer, and Haensel}}]{Chabanat:1997qh}
\bibinfo{author}{\bibfnamefont{E.}~\bibnamefont{Chabanat}},
  \bibinfo{author}{\bibfnamefont{J.}~\bibnamefont{Meyer}},
  \bibinfo{author}{\bibfnamefont{P.}~\bibnamefont{Bonche}},
  \bibinfo{author}{\bibfnamefont{R.}~\bibnamefont{Schaeffer}},
  \bibnamefont{and} \bibinfo{author}{\bibfnamefont{P.}~\bibnamefont{Haensel}},
  \bibinfo{journal}{Nucl. Phys.} \textbf{\bibinfo{volume}{A627}},
  \bibinfo{pages}{710} (\bibinfo{year}{1997}).

\bibitem[{\citenamefont{Chamel et~al.}(2009)\citenamefont{Chamel, Goriely, and
  Pearson}}]{Chamel:2009yx}
\bibinfo{author}{\bibfnamefont{N.}~\bibnamefont{Chamel}},
  \bibinfo{author}{\bibfnamefont{S.}~\bibnamefont{Goriely}}, \bibnamefont{and}
  \bibinfo{author}{\bibfnamefont{J.~M.} \bibnamefont{Pearson}},
  \bibinfo{journal}{Phys. Rev.} \textbf{\bibinfo{volume}{C80}},
  \bibinfo{pages}{065804} (\bibinfo{year}{2009}), \eprint{0911.3346}.

\bibitem[{\citenamefont{Zhang and Chen}(2016)}]{Zhang:2015vaa}
\bibinfo{author}{\bibfnamefont{Z.}~\bibnamefont{Zhang}} \bibnamefont{and}
  \bibinfo{author}{\bibfnamefont{L.-W.} \bibnamefont{Chen}},
  \bibinfo{journal}{Phys. Rev.} \textbf{\bibinfo{volume}{C94}},
  \bibinfo{pages}{064326} (\bibinfo{year}{2016}), \eprint{1510.06459}.

\bibitem[{\citenamefont{Zhou et~al.}(2019)\citenamefont{Zhou, Chen, and
  Zhang}}]{Zhou:2019omw}
\bibinfo{author}{\bibfnamefont{Y.}~\bibnamefont{Zhou}},
  \bibinfo{author}{\bibfnamefont{L.-W.} \bibnamefont{Chen}}, \bibnamefont{and}
  \bibinfo{author}{\bibfnamefont{Z.}~\bibnamefont{Zhang}},
  \bibinfo{journal}{Phys. Rev.} \textbf{\bibinfo{volume}{D99}},
  \bibinfo{pages}{121301} (\bibinfo{year}{2019}), \eprint{1901.11364}.

\bibitem[{\citenamefont{Chabanat et~al.}(1998)\citenamefont{Chabanat, Bonche,
  Haensel, Meyer, and Schaeffer}}]{Chabanat:1997un}
\bibinfo{author}{\bibfnamefont{E.}~\bibnamefont{Chabanat}},
  \bibinfo{author}{\bibfnamefont{P.}~\bibnamefont{Bonche}},
  \bibinfo{author}{\bibfnamefont{P.}~\bibnamefont{Haensel}},
  \bibinfo{author}{\bibfnamefont{J.}~\bibnamefont{Meyer}}, \bibnamefont{and}
  \bibinfo{author}{\bibfnamefont{R.}~\bibnamefont{Schaeffer}},
  \bibinfo{journal}{Nucl. Phys.} \textbf{\bibinfo{volume}{A635}},
  \bibinfo{pages}{231} (\bibinfo{year}{1998}), \bibinfo{note}{[Erratum: Nucl.
  Phys.A643,441(1998)]}.

\bibitem[{\citenamefont{Wang et~al.}(2016)\citenamefont{Wang, Friar, and
  Hayes}}]{Wang:2016rqh}
\bibinfo{author}{\bibfnamefont{X.~B.} \bibnamefont{Wang}},
  \bibinfo{author}{\bibfnamefont{J.~L.} \bibnamefont{Friar}}, \bibnamefont{and}
  \bibinfo{author}{\bibfnamefont{A.~C.} \bibnamefont{Hayes}},
  \bibinfo{journal}{Phys. Rev.} \textbf{\bibinfo{volume}{C94}},
  \bibinfo{pages}{034314} (\bibinfo{year}{2016}), \eprint{1607.02149}.

\bibitem[{\citenamefont{Wang and Gyulassy}(1991)}]{Wang:1991hta}
\bibinfo{author}{\bibfnamefont{X.-N.} \bibnamefont{Wang}} \bibnamefont{and}
  \bibinfo{author}{\bibfnamefont{M.}~\bibnamefont{Gyulassy}},
  \bibinfo{journal}{Phys. Rev.} \textbf{\bibinfo{volume}{D44}},
  \bibinfo{pages}{3501} (\bibinfo{year}{1991}).

\bibitem[{\citenamefont{Wang}(1997)}]{Wang:1996yf}
\bibinfo{author}{\bibfnamefont{X.-N.} \bibnamefont{Wang}},
  \bibinfo{journal}{Phys. Rept.} \textbf{\bibinfo{volume}{280}},
  \bibinfo{pages}{287} (\bibinfo{year}{1997}), \eprint{hep-ph/9605214}.

\bibitem[{\citenamefont{Andersson et~al.}(1983)\citenamefont{Andersson,
  Gustafson, Ingelman, and Sjostrand}}]{Andersson:1983ia}
\bibinfo{author}{\bibfnamefont{B.}~\bibnamefont{Andersson}},
  \bibinfo{author}{\bibfnamefont{G.}~\bibnamefont{Gustafson}},
  \bibinfo{author}{\bibfnamefont{G.}~\bibnamefont{Ingelman}}, \bibnamefont{and}
  \bibinfo{author}{\bibfnamefont{T.}~\bibnamefont{Sjostrand}},
  \bibinfo{journal}{Phys. Rept.} \textbf{\bibinfo{volume}{97}},
  \bibinfo{pages}{31} (\bibinfo{year}{1983}).

\bibitem[{\citenamefont{Zhang et~al.}(2000)\citenamefont{Zhang, Ko, Li, and
  Lin}}]{Zhang:1999bd}
\bibinfo{author}{\bibfnamefont{B.}~\bibnamefont{Zhang}},
  \bibinfo{author}{\bibfnamefont{C.}~\bibnamefont{Ko}},
  \bibinfo{author}{\bibfnamefont{B.-A.} \bibnamefont{Li}}, \bibnamefont{and}
  \bibinfo{author}{\bibfnamefont{Z.-W.} \bibnamefont{Lin}},
  \bibinfo{journal}{Phys. Rev.} \textbf{\bibinfo{volume}{C61}},
  \bibinfo{pages}{067901} (\bibinfo{year}{2000}), \eprint{nucl-th/9907017}.

\bibitem[{\citenamefont{Zhang}(1998)}]{Zhang:1997ej}
\bibinfo{author}{\bibfnamefont{B.}~\bibnamefont{Zhang}},
  \bibinfo{journal}{Comput. Phys. Commun.} \textbf{\bibinfo{volume}{109}},
  \bibinfo{pages}{193} (\bibinfo{year}{1998}), \eprint{nucl-th/9709009}.

\bibitem[{\citenamefont{Li and Ko}(1995)}]{Li:1995pra}
\bibinfo{author}{\bibfnamefont{B.-A.} \bibnamefont{Li}} \bibnamefont{and}
  \bibinfo{author}{\bibfnamefont{C.~M.} \bibnamefont{Ko}},
  \bibinfo{journal}{Phys. Rev.} \textbf{\bibinfo{volume}{C52}},
  \bibinfo{pages}{2037} (\bibinfo{year}{1995}), \eprint{nucl-th/9505016}.

\bibitem[{\citenamefont{Lin and Ko}(2002)}]{Lin:2001zk}
\bibinfo{author}{\bibfnamefont{Z.-W.} \bibnamefont{Lin}} \bibnamefont{and}
  \bibinfo{author}{\bibfnamefont{C.}~\bibnamefont{Ko}}, \bibinfo{journal}{Phys.
  Rev.} \textbf{\bibinfo{volume}{C65}}, \bibinfo{pages}{034904}
  (\bibinfo{year}{2002}), \eprint{nucl-th/0108039}.

\bibitem[{\citenamefont{Bass et~al.}(1998)}]{Bass:1998ca}
\bibinfo{author}{\bibfnamefont{S.~A.} \bibnamefont{Bass}} \bibnamefont{et~al.},
  \bibinfo{journal}{Prog. Part. Nucl. Phys.} \textbf{\bibinfo{volume}{41}},
  \bibinfo{pages}{255} (\bibinfo{year}{1998}), \eprint{nucl-th/9803035}.

\bibitem[{\citenamefont{Bleicher et~al.}(1999)}]{Bleicher:1999xi}
\bibinfo{author}{\bibfnamefont{M.}~\bibnamefont{Bleicher}}
  \bibnamefont{et~al.}, \bibinfo{journal}{J. Phys.}
  \textbf{\bibinfo{volume}{G25}}, \bibinfo{pages}{1859} (\bibinfo{year}{1999}),
  \eprint{hep-ph/9909407}.

\bibitem[{\citenamefont{Abelev et~al.}(2009)}]{Abelev:2008ab}
\bibinfo{author}{\bibfnamefont{B.~I.} \bibnamefont{Abelev}}
  \bibnamefont{et~al.} (\bibinfo{collaboration}{STAR}), \bibinfo{journal}{Phys.
  Rev.} \textbf{\bibinfo{volume}{C79}}, \bibinfo{pages}{034909}
  (\bibinfo{year}{2009}), \eprint{0808.2041}.

\bibitem[{\citenamefont{Kharzeev and Nardi}(2001)}]{Kharzeev:2000ph}
\bibinfo{author}{\bibfnamefont{D.}~\bibnamefont{Kharzeev}} \bibnamefont{and}
  \bibinfo{author}{\bibfnamefont{M.}~\bibnamefont{Nardi}},
  \bibinfo{journal}{Phys. Lett.} \textbf{\bibinfo{volume}{B507}},
  \bibinfo{pages}{121} (\bibinfo{year}{2001}), \eprint{nucl-th/0012025}.

\bibitem[{\citenamefont{Kharzeev et~al.}(2005)\citenamefont{Kharzeev, Levin,
  and Nardi}}]{Kharzeev:2001yq}
\bibinfo{author}{\bibfnamefont{D.}~\bibnamefont{Kharzeev}},
  \bibinfo{author}{\bibfnamefont{E.}~\bibnamefont{Levin}}, \bibnamefont{and}
  \bibinfo{author}{\bibfnamefont{M.}~\bibnamefont{Nardi}},
  \bibinfo{journal}{Phys. Rev.} \textbf{\bibinfo{volume}{C71}},
  \bibinfo{pages}{054903} (\bibinfo{year}{2005}), \eprint{hep-ph/0111315}.

\bibitem[{\citenamefont{Back et~al.}(2002)}]{Back:2002uc}
\bibinfo{author}{\bibfnamefont{B.~B.} \bibnamefont{Back}} \bibnamefont{et~al.}
  (\bibinfo{collaboration}{PHOBOS}), \bibinfo{journal}{Phys. Rev.}
  \textbf{\bibinfo{volume}{C65}}, \bibinfo{pages}{061901}
  (\bibinfo{year}{2002}), \eprint{nucl-ex/0201005}.

\bibitem[{\citenamefont{Alver et~al.}(2007)}]{Alver:2006wh}
\bibinfo{author}{\bibfnamefont{B.}~\bibnamefont{Alver}} \bibnamefont{et~al.}
  (\bibinfo{collaboration}{PHOBOS}), \bibinfo{journal}{Phys. Rev. Lett.}
  \textbf{\bibinfo{volume}{98}}, \bibinfo{pages}{242302}
  (\bibinfo{year}{2007}), \eprint{nucl-ex/0610037}.

\bibitem[{\citenamefont{Miller et~al.}(2007)\citenamefont{Miller, Reygers,
  Sanders, and Steinberg}}]{Miller:2007ri}
\bibinfo{author}{\bibfnamefont{M.~L.} \bibnamefont{Miller}},
  \bibinfo{author}{\bibfnamefont{K.}~\bibnamefont{Reygers}},
  \bibinfo{author}{\bibfnamefont{S.~J.} \bibnamefont{Sanders}},
  \bibnamefont{and}
  \bibinfo{author}{\bibfnamefont{P.}~\bibnamefont{Steinberg}},
  \bibinfo{journal}{Ann. Rev. Nucl. Part. Sci.} \textbf{\bibinfo{volume}{57}},
  \bibinfo{pages}{205} (\bibinfo{year}{2007}), \eprint{nucl-ex/0701025}.

\bibitem[{\citenamefont{Xu et~al.}(2014)\citenamefont{Xu, Pang, and
  Wang}}]{Xu:2014ada}
\bibinfo{author}{\bibfnamefont{H.-j.} \bibnamefont{Xu}},
  \bibinfo{author}{\bibfnamefont{L.}~\bibnamefont{Pang}}, \bibnamefont{and}
  \bibinfo{author}{\bibfnamefont{Q.}~\bibnamefont{Wang}},
  \bibinfo{journal}{Phys. Rev.} \textbf{\bibinfo{volume}{C89}},
  \bibinfo{pages}{064902} (\bibinfo{year}{2014}), \eprint{1404.2663}.

\bibitem[{\citenamefont{Aamodt et~al.}(2011)}]{Aamodt:2010cz}
\bibinfo{author}{\bibfnamefont{K.}~\bibnamefont{Aamodt}} \bibnamefont{et~al.}
  (\bibinfo{collaboration}{ALICE}), \bibinfo{journal}{Phys. Rev. Lett.}
  \textbf{\bibinfo{volume}{106}}, \bibinfo{pages}{032301}
  (\bibinfo{year}{2011}), \eprint{1012.1657}.

\bibitem[{\citenamefont{Adare et~al.}(2016)}]{Adare:2015bua}
\bibinfo{author}{\bibfnamefont{A.}~\bibnamefont{Adare}} \bibnamefont{et~al.}
  (\bibinfo{collaboration}{PHENIX}), \bibinfo{journal}{Phys. Rev. C}
  \textbf{\bibinfo{volume}{93}}, \bibinfo{pages}{024901}
  (\bibinfo{year}{2016}), \eprint{1509.06727}.

\bibitem[{\citenamefont{Adare et~al.}(2018)}]{Adare:2018toe}
\bibinfo{author}{\bibfnamefont{A.}~\bibnamefont{Adare}} \bibnamefont{et~al.}
  (\bibinfo{collaboration}{PHENIX}), \bibinfo{journal}{Phys. Rev. Lett.}
  \textbf{\bibinfo{volume}{121}}, \bibinfo{pages}{222301}
  (\bibinfo{year}{2018}), \eprint{1807.11928}.

\bibitem[{\citenamefont{Wang and Sorensen}(2014)}]{Wang:2014qxa}
\bibinfo{author}{\bibfnamefont{H.}~\bibnamefont{Wang}} \bibnamefont{and}
  \bibinfo{author}{\bibfnamefont{P.}~\bibnamefont{Sorensen}}
  (\bibinfo{collaboration}{STAR}), \bibinfo{journal}{Nucl. Phys. A}
  \textbf{\bibinfo{volume}{932}}, \bibinfo{pages}{169} (\bibinfo{year}{2014}),
  \eprint{1406.7522}.

\bibitem[{\citenamefont{Moreland et~al.}(2015)\citenamefont{Moreland, Bernhard,
  and Bass}}]{Moreland:2014oya}
\bibinfo{author}{\bibfnamefont{J.~S.} \bibnamefont{Moreland}},
  \bibinfo{author}{\bibfnamefont{J.~E.} \bibnamefont{Bernhard}},
  \bibnamefont{and} \bibinfo{author}{\bibfnamefont{S.~A.} \bibnamefont{Bass}},
  \bibinfo{journal}{Phys.Rev.} \textbf{\bibinfo{volume}{C92}},
  \bibinfo{pages}{011901} (\bibinfo{year}{2015}), \eprint{1412.4708}.

\bibitem[{\citenamefont{Bernhard et~al.}(2016)\citenamefont{Bernhard, Moreland,
  Bass, Liu, and Heinz}}]{Bernhard:2016tnd}
\bibinfo{author}{\bibfnamefont{J.~E.} \bibnamefont{Bernhard}},
  \bibinfo{author}{\bibfnamefont{J.~S.} \bibnamefont{Moreland}},
  \bibinfo{author}{\bibfnamefont{S.~A.} \bibnamefont{Bass}},
  \bibinfo{author}{\bibfnamefont{J.}~\bibnamefont{Liu}}, \bibnamefont{and}
  \bibinfo{author}{\bibfnamefont{U.}~\bibnamefont{Heinz}},
  \bibinfo{journal}{Phys. Rev.} \textbf{\bibinfo{volume}{C94}},
  \bibinfo{pages}{024907} (\bibinfo{year}{2016}), \eprint{1605.03954}.

\bibitem[{\citenamefont{Goldschmidt et~al.}(2015)\citenamefont{Goldschmidt,
  Qiu, Shen, and Heinz}}]{Goldschmidt:2015kpa}
\bibinfo{author}{\bibfnamefont{A.}~\bibnamefont{Goldschmidt}},
  \bibinfo{author}{\bibfnamefont{Z.}~\bibnamefont{Qiu}},
  \bibinfo{author}{\bibfnamefont{C.}~\bibnamefont{Shen}}, \bibnamefont{and}
  \bibinfo{author}{\bibfnamefont{U.}~\bibnamefont{Heinz}},
  \bibinfo{journal}{Phys. Rev.} \textbf{\bibinfo{volume}{C92}},
  \bibinfo{pages}{044903} (\bibinfo{year}{2015}), \eprint{1507.03910}.

\bibitem[{\citenamefont{Giacalone}(2019)}]{Giacalone:2018apa}
\bibinfo{author}{\bibfnamefont{G.}~\bibnamefont{Giacalone}},
  \bibinfo{journal}{Phys. Rev.} \textbf{\bibinfo{volume}{C99}},
  \bibinfo{pages}{024910} (\bibinfo{year}{2019}), \eprint{1811.03959}.

\bibitem[{\citenamefont{Pang et~al.}(2019)\citenamefont{Pang, Zhou, and
  Wang}}]{Pang:2019aqb}
\bibinfo{author}{\bibfnamefont{L.-G.} \bibnamefont{Pang}},
  \bibinfo{author}{\bibfnamefont{K.}~\bibnamefont{Zhou}}, \bibnamefont{and}
  \bibinfo{author}{\bibfnamefont{X.-N.} \bibnamefont{Wang}}
  (\bibinfo{year}{2019}), \eprint{1906.06429}.

\bibitem[{\citenamefont{Giacalone}(2020)}]{Giacalone:2019pca}
\bibinfo{author}{\bibfnamefont{G.}~\bibnamefont{Giacalone}},
  \bibinfo{journal}{Phys. Rev. Lett.} \textbf{\bibinfo{volume}{124}},
  \bibinfo{pages}{202301} (\bibinfo{year}{2020}), \eprint{1910.04673}.

\bibitem[{\citenamefont{Xiao et~al.}(2009)\citenamefont{Xiao, Li, Chen, Yong,
  and Zhang}}]{Xiao:2008vm}
\bibinfo{author}{\bibfnamefont{Z.}~\bibnamefont{Xiao}},
  \bibinfo{author}{\bibfnamefont{B.-A.} \bibnamefont{Li}},
  \bibinfo{author}{\bibfnamefont{L.-W.} \bibnamefont{Chen}},
  \bibinfo{author}{\bibfnamefont{G.-C.} \bibnamefont{Yong}}, \bibnamefont{and}
  \bibinfo{author}{\bibfnamefont{M.}~\bibnamefont{Zhang}},
  \bibinfo{journal}{Phys. Rev. Lett.} \textbf{\bibinfo{volume}{102}},
  \bibinfo{pages}{062502} (\bibinfo{year}{2009}), \eprint{0808.0186}.

\bibitem[{\citenamefont{Wei et~al.}(2014)\citenamefont{Wei, Li, Xu, and
  Chen}}]{Wei:2013sfa}
\bibinfo{author}{\bibfnamefont{G.-F.} \bibnamefont{Wei}},
  \bibinfo{author}{\bibfnamefont{B.-A.} \bibnamefont{Li}},
  \bibinfo{author}{\bibfnamefont{J.}~\bibnamefont{Xu}}, \bibnamefont{and}
  \bibinfo{author}{\bibfnamefont{L.-W.} \bibnamefont{Chen}},
  \bibinfo{journal}{Phys. Rev.} \textbf{\bibinfo{volume}{C90}},
  \bibinfo{pages}{014610} (\bibinfo{year}{2014}), \eprint{1309.7717}.

\bibitem[{\citenamefont{Hartnack et~al.}(2018)\citenamefont{Hartnack, Le~Fevre,
  Leifels, and Aichelin}}]{Hartnack:2018sih}
\bibinfo{author}{\bibfnamefont{C.}~\bibnamefont{Hartnack}},
  \bibinfo{author}{\bibfnamefont{A.}~\bibnamefont{Le~Fevre}},
  \bibinfo{author}{\bibfnamefont{Y.}~\bibnamefont{Leifels}}, \bibnamefont{and}
  \bibinfo{author}{\bibfnamefont{J.}~\bibnamefont{Aichelin}}
  (\bibinfo{year}{2018}), \eprint{1808.09868}.

\bibitem[{\citenamefont{Helenius et~al.}(2017)\citenamefont{Helenius,
  Paukkunen, and Eskola}}]{Helenius:2016dsk}
\bibinfo{author}{\bibfnamefont{I.}~\bibnamefont{Helenius}},
  \bibinfo{author}{\bibfnamefont{H.}~\bibnamefont{Paukkunen}},
  \bibnamefont{and} \bibinfo{author}{\bibfnamefont{K.~J.}
  \bibnamefont{Eskola}}, \bibinfo{journal}{Eur. Phys. J.}
  \textbf{\bibinfo{volume}{C77}}, \bibinfo{pages}{148} (\bibinfo{year}{2017}),
  \eprint{1606.06910}.

\end{thebibliography}

\end{document}